\documentclass[prl,aps,twocolumn,superscriptaddress]{revtex4-2}
\usepackage{graphicx,color,setspace}
\usepackage[utf8]{inputenc}
\usepackage{mathtools}
\usepackage[normalem]{ulem}
\usepackage{tabularx}
\usepackage{enumerate}%
\usepackage{float}
\usepackage[normalem]{ulem}
\usepackage{comment}
\usepackage{soul}
\usepackage[caption=false]{subfig} %
\usepackage{ragged2e} %
\DeclareCaptionJustification{justified}{\justifying}

\usepackage{pifont}
\let\oldding\ding%
\renewcommand{\ding}[2][1]{\scalebox{#1}{\oldding{#2}}}%

\usepackage[labelfont=bf]{caption}

\begin{document}

\author{Camilla Beneduce}
\affiliation{Dipartimento di Fisica, Sapienza Universit\`{a} di Roma, P.le Aldo Moro 5, 00185 Rome, Italy}
\author{Diogo E. P. Pinto}
\affiliation{Dipartimento di Fisica, Sapienza Universit\`{a} di Roma, P.le Aldo Moro 5, 00185 Rome, Italy}
\author{Lorenzo Rovigatti}
\affiliation{Dipartimento di Fisica, Sapienza Universit\`{a} di Roma, P.le Aldo Moro 5, 00185 Rome, Italy}
\author{Flavio Romano}
\affiliation{Department of Molecular Sciences and Nanosystems, Ca' Foscari University of Venice, Via Torino 155, 30171 Venezia-Mestre, Italy}
\author{Petr \v{S}ulc}
\affiliation{School of Molecular Sciences and Center for Molecular Design and Biomimetics, The Biodesign Institute, Arizona State University, 1001 South McAllister Avenue, Tempe, Arizona 85281, USA}
\affiliation{Center for Biological Physics, Arizona State University, Tempe, Arizona 85281, USA}
\affiliation{TU Munich, School of Natural Sciences, Department of Bioscience, Garching, Germany}
\author{Francesco Sciortino}
\affiliation{Dipartimento di Fisica, Sapienza Universit\`{a} di Roma, P.le Aldo Moro 5, 00185 Rome, Italy}
\author{John Russo}
\affiliation{Dipartimento di Fisica, Sapienza Universit\`{a} di Roma, P.le Aldo Moro 5, 00185 Rome, Italy}

\title{A Falsifiability Test for Classical Nucleation Theory}

\begin{abstract}
\noindent Classical nucleation theory (CNT) is built upon the capillarity approximation, \textit{i.e.}, the assumption that the nucleation properties can be inferred from the bulk properties of the melt and the crystal.
Although CNT's simplicity and usefulness cannot be overstated, experiments and simulations regularly uncover significant deviations from its predictions, which are often reconciled through phenomenological extensions of the CNT, fueling the debate over the general validity of the theory.
In this study, we present a falsifiability test for any nucleation theory grounded in the capillarity approximation.
We focus on cases where the theory predicts no differences in nucleation rates between different crystal polymorphs. We then introduce a system in which all polymorphs have the same free energy (both bulk and interfacial) across all state points. Through extensive molecular simulations, we show that the polymorphs exhibit remarkably different nucleation properties, directly contradicting CNT’s predictions.
We argue that CNT’s primary limitation lies in its neglect of structural fluctuations within the liquid phase.

\end{abstract}

\maketitle

Citing Baron Peters (who in turn paraphrases Popper~\cite{popper1963conjectures}): ``The most convincing test of a theory comes from special cases where the theory should fail if it is not true''~\cite{peters2022crystal}. Here, we present a system that is ideally suited to test one of the basic assumptions of Classical Nucleation Theory (CNT), \textit{i.e.} the so-called \emph{capillarity approximation}.
The capillarity approximation assumes that the critical nucleus has the same thermodynamic properties (surface tension, density, etc.) of the bulk phase, which allows for a simple calculation of the free-energy barrier for the phase transition. 
CNT defines the nucleation rate as $K e{^{-\Delta G(n_c)/k_B T}}$, where $K$ is the kinetic prefactor that accounts for the attachment rate of particles to the nucleus. $\Delta G(n_c)$ is the free-energy barrier for nucleation and is obtained by maximising $\Delta G(n)=-n|\Delta\mu|+\alpha n^{2/3} \gamma $, \textit{i.e.} the Gibbs free energy cost to form a nucleus of $n$ particles at constant pressure and temperature, where
$\Delta\mu$ is the chemical potential difference between the crystal and the melt, $\gamma$ is the interfacial free energy, and $\alpha$ is a proportionality constant accounting for the shape of the nucleus~\cite{kelton2010nucleation}.
Due to its relative simplicity and predictive power, CNT has been perhaps the most used theoretical model to describe nucleation processes. Although it has been successful in many cases ~\cite{miller1983homogeneous, strey1994problem, manka2010homogeneous, prado2019successful}, there are also numerous cases where its predictions do not align with experimental and simulation results~\cite{lihavainen2001homogeneous, zhang2007does, iland2007argon, erdemir2009nucleation}. In such instances, various phenomenological extensions to the theory~\cite{tolman1949effect, langer1968theory, plischke1984active, rikvold1994metastable, ford1997nucleation, ramos1999test, sides1999kinetic, reguera2004extended, merikanto2007origin, gebauer2011prenucleation, prestipino2012systematic,espinosa2016seeding, coli2021artificial, gispen2023brute, gispen2024finding} have been proposed to reconcile these discrepancies without abandoning the capillarity approximation, for example by introducing temperature-dependent interfacial free energies.

Usual tests of CNT consists in comparing the measured nucleation rates with theoretical values.
The problem with this approach is that the results depend strongly on quantities, notably $\gamma$, which are state-dependent and very difficult to measure accurately at conditions where homogeneous nucleation occurs. Citing Oxtoby, ``Nucleation theory is one of the few areas of science in which agreement of predicted and measured rates to within several orders of magnitude is considered a major success''~\cite{oxtoby1998nucleation}.

Here we present a falsifiability test not only for CNT but also for any nucleation theory grounded
in the capillarity approximation.
Instead of accurately predicting nucleation rates, we focus on the polymorphic composition of the crystalline phase~\cite{ten1999homogeneous,piaggi2018kinetcting}, i.e. the ability of a material to exist in more than one crystalline structure~\cite{van2018molecular,burcham2024pharmaceutical,ostwald1897studien,de2016nucleation,sadigh2021metastable,bupathy2022temperature,sleutel2014observing}.
We introduce a binary mixture with three polymorphs that possess identical free energies (both bulk and interfacial) at all state points.
The three polymorphs are isotypic forms of the cubic diamond crystal, \textit{i.e.} they share the same atomic positions and symmetry of the crystal lattice, but differ in the way the different species are arranged on the lattice sites.
Within the capillarity approximation all these structures should have identical nucleation properties, given that there are no free-energy differences between the different polymorphs and that they all form from the same liquid phase. Instead, via molecular simulations, we find that the nucleation properties of the three polymorphs are radically different. Interestingly, we find that the polymorph that nucleates more easily is the one with the largest unit cell, and the one that nucleates the least is the one with the smallest unit cell.
To account for the difference in nucleation properties between the polymorphs, we show that the orientational order of the melt is closest to the polymorph that nucleates more frequently.

\begin{figure}[t]
    \centering
    \includegraphics[width=0.45\textwidth]{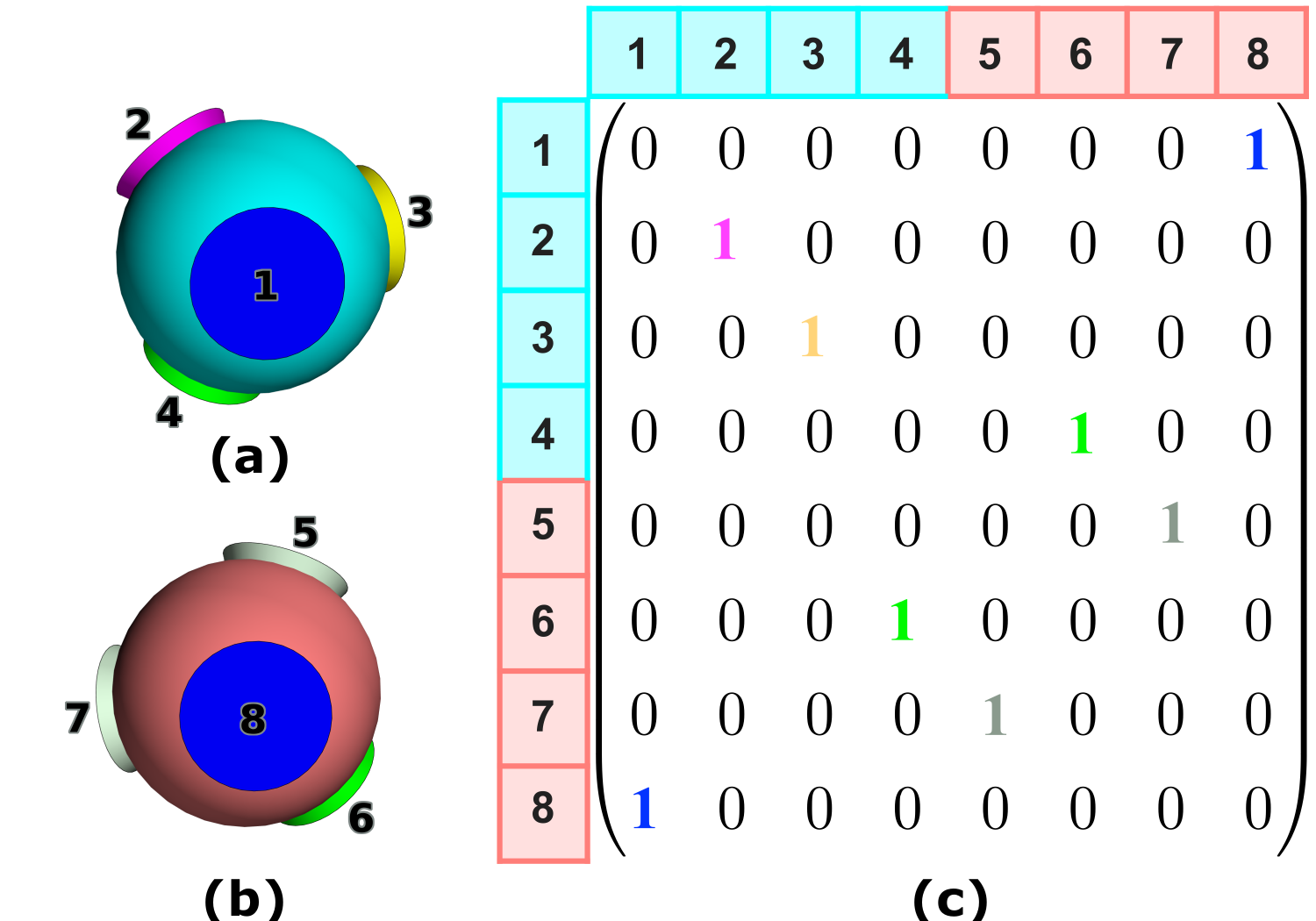}
    \caption{\textbf{N2c8 design.} Binary mixture of patchy particles SAT-designed to exclusively self-assemble into a cubic diamond crystal. The two species, depicted in cyan (a) and red (b), have four patches (numbered from $1$ to $8$) tetrahedrally arranged that bind according to the interaction matrix (c) where the ones indicate the interacting patches. Matching colors appear for complementary patches (off-diagonal ones), while unique colors represent self-complementary interactions (diagonal ones).}
\label{fig:N2c8}
\end{figure}

\begin{figure*}[t]
    \centering
    \includegraphics[width=0.99\textwidth]{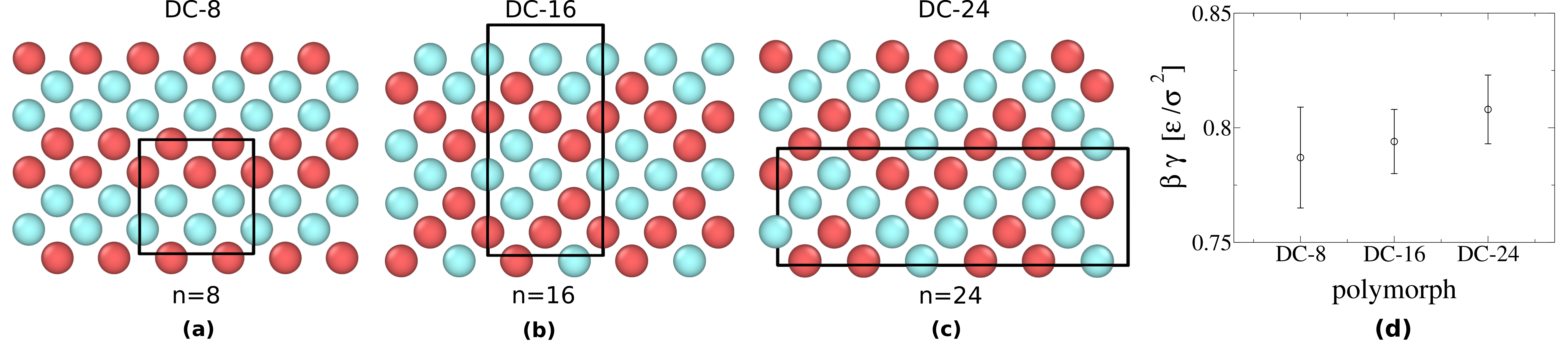}
\caption{\textbf{Overview of cubic diamond polymorphs.} The SAT-assembly framework allows to enumerate all the possible arrangements of the N2c8 particles in the 48 particle cubic diamond lattice. Three periodic patterns are identified: DC-8 (a), DC-16 (b), and DC-24 (c). They are labeled and displayed in increasing order of unit cell size (black box). In (d) their surface tensions are plotted. Each estimate comes from an average of four independent Monte Carlo simulations in the grand canonical ensemble. All values are considered equal within an error of $3\%$.}
\label{fig:poly}
\end{figure*}

The model system is presented in Fig.~\ref{fig:N2c8}. It is a binary mixture of tetravalent patchy particles, \textit{i.e.} particles that have a hard-core repulsion and attractive spots tetrahedrally located on their surface, as detailed in Supplemental Material~\cite{supp}. The interaction between the patches is specific and defined by the interaction matrix in Fig. 1c, whose elements $m_{ij}$ are equal to $1$ only if patches $i$ and $j$ can bind.
The specificity of interactions in patchy particles model systems can be experimentally realized exploiting the predictable and controllable interactions (Watson-Crick base pairing) of DNA~\cite{jaekel2020insights,li2024dna,zhou2024colloidal}, with one of the most promising approaches being the use of DNA origami~\cite{cumberworth2022simulations, doye2023oxdna}. This technique has recently been successfully applied to the self-assembly of a pyrochlore lattice, confirming the feasibility of the approach~\cite{liu2024inverse}.
The design in Fig.~\ref{fig:N2c8}, called N2c8, since it uses two species and eight different patch types, was originally introduced in the context of SAT-assembly~\cite{russo2022sat,romano2020designing} as a system able to self-assemble exclusively into the cubic diamond structure (DC) while avoiding the hexagonal diamond one~\cite{rovigatti2022simple}.
As explained in the dedicated section in Supplemental Material~\cite{supp}, the SAT-assembly algorithm can also be used to list all possible ways to fill the lattice positions of the target cell by particles belonging to a selected design (in our case N2c8).
This allows us to automatically identify all those possible polymorphs that, regardless of their nucleation abilities, are compatible with a cubic diamond lattice of a certain size (defying McCrone's law~\footnote{the number of forms known for a given compound is proportional to the time and money spent in research on that compound}).
We find that there are three possible periodically repeated patterns of N2c8 patchy particles within the lattice positions of a 48-particle cubic diamond cell. All these arrangements are illustrated in Fig.~\ref{fig:poly} where we display the [001] plane. 
The three structures belong to the P1 space group, meaning that there is no symmetry other than the translational one. Yet, they differ in the unit cell size (black box in Fig.~\ref{fig:poly}), which we use for labeling: DC-X where $X\in [8,16,24]$ refers to the size of the unit cell of each polymorph. We use here the word unit cell to indicate the smallest repeating unit of the two species, regardless of patch coloring, that, when stacked together, creates the crystal lattice, as shown in Fig.~\ref{fig:poly}. It is worth noting that if the patch arrangement in the unit cell is also considered (not shown in Fig.~\ref{fig:poly}), the DC-8, DC-16, DC-24 have unit cells of 16, 16, and 48 atoms, respectively. In the following we will indicate the size of the unit cell only referring to the species occupation, regardless of the patch arrangement.
Although other polymorphs with larger unit cell sizes exist, they do not appear in our simulations because nucleation involves critical nuclei smaller than 30 particles under the conditions investigated. We speculate that crystals with unit cells larger than the critical nucleus size $n_c$, which is temperature-dependent, are prevented from nucleating.

\noindent The three polymorphs investigated here are isotypic, their composition is always equimolar and they have only translational symmetry.
In each of the crystals the same number and type of bonds are established, and lattice vibrations are controlled by the geometry of the patches, which is the same for both species of the mixture. Therefore, all polymorphs have the same bulk free energy. Moreover, we verify the equality of their (solid/liquid) interfacial free energies. With successive umbrella sampling simulations ~\cite{virnau2004calculation} we have computed the free energy cost of forming an interface between each polymorph and the same liquid phase at coexistence conditions (determined via direct coexistence simulations). The resulting average values of surface free energy are reported in Fig. 2d for the [100] square plane. We leave the full description on how the computation is performed in the dedicated section of the Supplemental Material~\cite{supp}. Here we just emphasize that the surface tensions of the three polymorphs are the same within the error (approximately $3\%$). The relevant macroscopic properties on which CNT is based are therefore the same in the DC-8, the DC-16 and the DC-24 structures.

Having characterized the bulk properties of the three polymorphs, we now consider their nucleation properties.
The N2c8 binary mixture has an azeotropic point at equimolar concentration~\cite{beneduce2023engineering}, meaning that an equimolar mixture will retain its composition during liquid-gas phase separation. This was shown to aid the nucleation process~\cite{beneduce2023two}, since crystal nucleation can occur in liquid droplets that have the same composition as the final crystalline structure.
We choose two state points where crystallisation was found to be favourable and we run extensive Monte Carlo simulations in the canonical ensemble to collect nucleation events. In the following, temperature $T$ is in unit of $\epsilon/k_{B}$, density $\rho$ is in unit of $1/\sigma^{3}$ and pressure $P$ in unit of $\epsilon/\sigma^{3}$, where $\sigma$ is the patchy particle diameter, $\epsilon$ is the square-well potential depth (see Supplemental Material~\cite{supp}) and $k_{B}=1$. In particular, we simulate $N=500$ patchy particles~\cite{rovigatti2018simulate} at equimolar concentration in the canonical ensemble at $T=0.1$, $\rho=0.35$ ($600$ trajectories, both with and without aggregation-volume bias (AVB) moves~\cite{rovigatti2018simulate, chen2000novel}) and at $T=0.104$, $\rho=0.4$ ($300$ trajectories with AVB dynamics). We label successfully nucleated those trajectories having a fraction of particles in the cubic diamond phase greater than $0.5$ and classify the obtained crystals. We use the total coherence, an order parameter based on spherical harmonics, to distinguish between liquid and crystalline particles  ~\cite{steinhardt1983bond,rein1996numerical,lechner2008accurate,desgranges2008crystallization}, and we implement a new order parameter (bond geometry and particle orientations must be taken into account) to classify the different polymorphs.
A full description of the order parameters can be found in the Section ``Total coherence order parameter'' in the Supplemental Material~\cite{supp}. As shown in Fig. 3a, the three polymorphs have different nucleation rates: the majority of nucleating trajectories form the DC-24 polymorph, the one with the largest unit cell size, while only a single nucleation event is observed for the one with the smallest unit cell size, the DC-8 polymorph. Specifically, for the $T=0.1$, $\rho=0.35$ state point, out of 600 trajectories, 43 form the DC-24, 15 the DC-16 and none the DC-8 with AVB dynamics and 30 form the DC-24, 11 the DC-16 and 1 the DC-8 with no AVB moves. Similar results are observed for the $T=0.104$ , $\rho=0.4$ state point (35, 6 and 0, respectively, out of 300 trajectories). In Fig. 3b we report the progress in time of the nucleus size for two typical trajectories: the blue one spontaneously self-assembling a DC-24 polymorph, and the green one crystallising into a DC-16 structure. Snapshots of the critical nucleus are also displayed, showing that the different polymorphs are already distinguishable at critical sizes. Additionally, we analyse the polymorphs growth (see Supplemental Material~\cite{supp}) to explicitly exclude the presence of cross nucleation of one polymorph on top of another.

\begin{figure*}[t]
    \centering
    \includegraphics[width=0.99\textwidth]{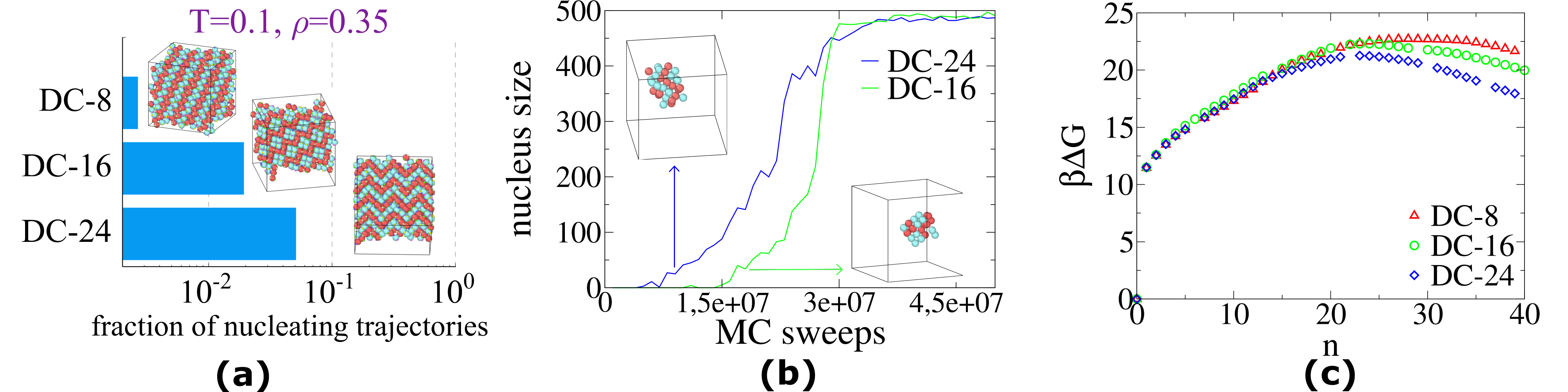}
\caption{\textbf{Direct nucleation simulations and nucleation barriers} (a) Fraction of trajectories ending up into each polymorph out of 600 with snapshots of final configurations. %
Simulations are run in the canonical ensemble at $T = 0.1$, $\rho= 0.35$, $N = 500$.
Interestingly, the polymorph with the largest unit cell size is the one that nucleates most frequently. (b) Time evolution of the number of crystalline particles with a snapshot of the critical nucleus for the DC-24 (blue) and the DC-16 (green) structures. (c) Gibbs free energy for the formation of crystal nucleus of size $n$ for the three polymorphs computed by means of umbrella sampling technique at $T=0.104$ and $P=0.018$. 
The different trend in nucleation simulations is confirmed: the smallest unit cell size polymorph has the highest barrier.
}
\label{fig:nuc_traj}
\end{figure*}

\noindent The different nucleation rates of the three polymorphs are reflected by the height of their free energy barriers, which we compute with the Umbrella Sampling technique (a brief description is provided in the Supplemental Material~\cite{supp}). For this, we run $NPT$ Monte Carlo simulations at $T=0.104$ and $P=0.018$ with a harmonic bias potential.
Each simulation is prepared by inserting in the liquid phase a crystalline seed of size $n_0$, where $n_0$ increases by 5 particles in successive simulations. The initial seeds have a roughly spherical shape and the same density for each polymorph. We carefully verify that the chosen bias potential and thermodynamic conditions ensure a good sampling, \textit{i.e.} that there is an appreciable overlap between simulations with successive $n_0$, that no spontaneous nucleation occurs, and that there is no change in polymorph identity during the simulation. The resulting barriers are reported in Fig. 3c. In order of increasing barrier height, we find DC-24, then DC-16, and finally DC-8, confirming that the polymorph with the lowest barrier is the one with the largest unit cell. The height difference between the DC-24 and DC-16 barrier is approximately $1 k_{B}T$, a value that aligns well with the fractions of trajectories nucleating into the two polymorphs. The same comparison cannot be made for the DC-8 case as only a single nucleation event is observed for this polymorph. The critical size $n_c$ for all nuclei is $n_c\lesssim 30$, \textit{i.e.} smaller than the largest unit cell (48 particles) used in our exhaustive search of competing polymorphs.

\begin{figure}[t]
    \centering
    \includegraphics[width=0.5\textwidth]{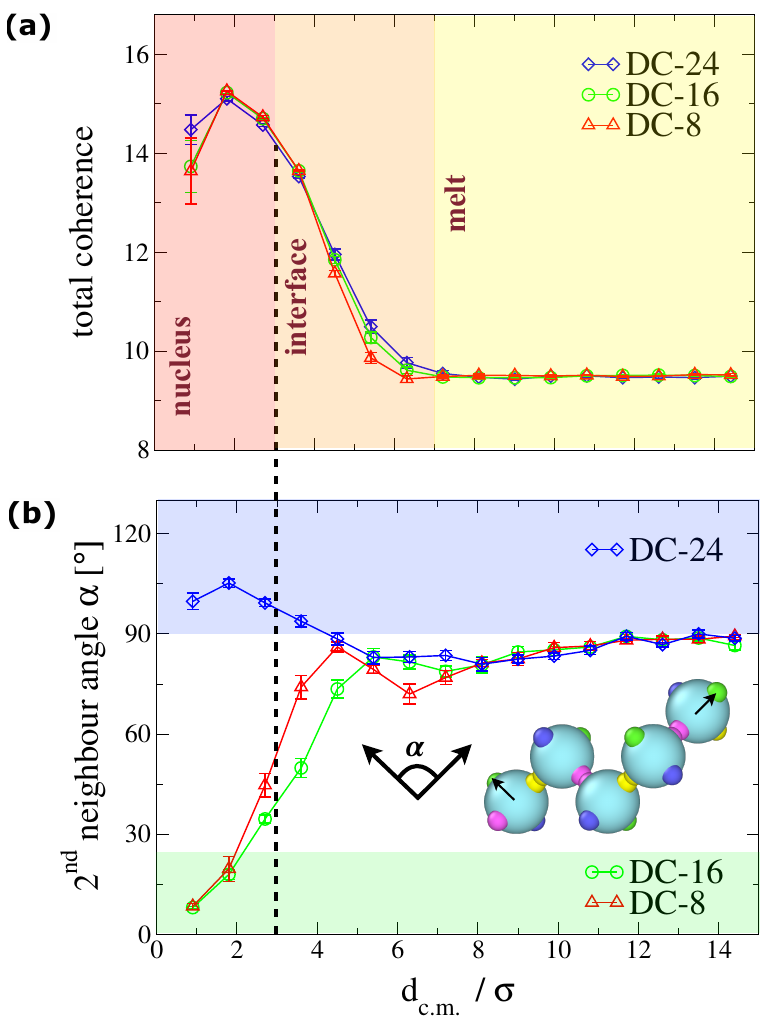}
    \caption{\textbf{Bond orientational order in the liquid phase.} Total coherence (a) and $\alpha$ angle (b) as a function of the distance of each particle from the center of mass $d_{c.m.}$ of the largest crystalline cluster. Points are computed by averaging on different configurations and trajectories characterised by a crystalline nucleus of  50 particles: a DC-24 (blue diamonds), a DC-16 (green circles), and a DC-8 (red triangles) nucleus. $\alpha$ defines the relative orientation between second neighbours of the same species as illustrated for patchy particles of the first species of the DC-24 polymorph in the inset of figure (b). The coloured bands in (a) helps locating the different regions: the nucleus, the interface and the melt. The coloured bands in (b) define the range of $\alpha$ values typical of the bulk DC-24 crystal (blue band) and of the bulk of both DC-16 and DC-8 (green band) polymorphs;  particles in the interfacial region as well as the ones in the melt have $\alpha$ angles characteristic of the DC-24 structure.
    }
    \label{liquid_or}
\end{figure}

We have observed that the three polymorphs exhibit different nucleation properties, specifically in terms of nucleation rates and barrier heights, and that these differences cannot be attributed to variations in the bulk properties of the crystals.
In the following, we investigate whether the different nucleation properties of the polymorphs can instead be traced back to the structural fluctuations within the melt.
We follow the idea that if the liquid phase from which the nucleus arises already exhibits some degree of order in particle orientation, this will favor the nucleation of the polymorph that is structurally closer to the melt~\cite{russo2012microscopic}. The three polymorphs, DC-8, DC-16, and DC-24, differ in fact by the orientation between neighbors of the same species. In particular, for each patchy particle, we can consider the angle $\alpha$ formed between the patch orientations of its two second-nearest neighbour of the same species (see inset of Fig.~\ref{liquid_or}b): for the bulk polymorphs we have $\alpha_\text{DC-24}\sim 109^\circ$, $\alpha_\text{DC-16}\sim 9^\circ$, and $\alpha_\text{DC-8}\sim 9^\circ$. In Fig.~\ref{liquid_or}b we plot the radial profile of $\alpha$ from the center of mass of nuclei of DC-24 (blue diamonds), DC-16 (green circles) and DC-8 (red triangles) polymorphs. We use configurations and trajectories from the umbrella sampling window of size $n\sim 50$ and, in order to compute $\alpha$, we consider only particles having two second nearest neighbours of the same species. To show the crystalline profile, in Fig.~\ref{liquid_or}a, we show the radial distribution of the total coherence which has high values (around 14) for the bulk diamond crystal, and small values (around 9.5) for the melt, irrespective of the polymorph. The figure shows the transition from the core of the nuclei at short distances to the melt at large distances.
In the core region, $\alpha$ assumes the value of the corresponding bulk polymorph ($\alpha_\text{DC-24}\sim 109^\circ$,  $\alpha_\text{DC-16}\sim 9^\circ$, and $\alpha_\text{DC-8}\sim 9^\circ$). In the interfacial region, instead, the value of $\alpha$ is always that of DC-24 polymorph, irrespectively of the nucleus type.
This shows that the orientational order in the melt resembles more closely that of the DC-24 polymorph, \textit{i.e.}, the polymorph which nucleates more frequently and has the lowest nucleation barrier. A similar argument can be made to rationalize why the DC-8 is the less frequently nucleating polymorph, as discussed in the Supplemental Material~\cite{supp}. Furthermore, a thorough analysis confirms that the observed difference in the nucleation frequency between the polymorphs cannot be attributed to the presence of favorable or unfavorable bonding sites on their surfaces (see Supplemental Material~\cite{supp}).

In conclusion,
unlike previous tests of classical nucleation theory (CNT), which focus on accurately measuring nucleation rates and comparing them to theoretical predictions, leading to a pass or fail outcome depending on the system and/or state point, in this work we propose an alternative approach. We examine cases where CNT predicts no difference in nucleation rates between different polymorphs. This shift in focus sidesteps the common reliance on precise free-energy measurements and their dependence on specific state points. Instead, it only requires ranking the polymorphs according to their nucleation frequency. Crucially, while discrepancies between CNT and measured nucleation rates can often be accounted for through \textit{ad hoc} extensions to the theory, our test cannot be satisfied by any modification of CNT built upon the capillarity approximation.
To run the falsifiability test
we introduce a binary mixture of patchy particles where three different polymorphs, despite having identical bulk and interfacial free energies, exhibit significantly different nucleation rates.

One of the shortcomings of CNT is its failure in accounting for the short-range order possessed by the supercooled liquid state~\cite{russo2012microscopic}. 
In our system, we have shown that the melt exhibits a local orientational order typical of the polymorph with the highest nucleation rate. This suggests that it is the structural fluctuations in the melt, both in terms of their size and orientational order, rather than the bulk properties of the infinitely large crystals, that determine which polymorph will nucleate. These arguments offer support to approaches beyond CNT that take into account the structural properties of the liquid phase~\cite{russo2012microscopic,russo2012selection,lutsko2024microscopic,whitelam2015statistical,zhang2017nonclassical,van2018molecular,russo2016crystal,james2019phase,lutsko2019crystals,becker2022unsupervised,schoonen2022crystal,gispen2023crystal}.

\section{Acknowledgments}

\begin{acknowledgments}
C.B., D.P., L.R., F.S. and J.R. acknowledge support by ICSC – Centro Nazionale di Ricerca in High Performance Computing, Big Data and Quantum Computing, funded by European Union – NextGenerationEU, and CINECA-ISCRA for HPC resources. 
D.P. and  F.S. acknowledge  support from MIUR-PRIN Grant No. 2022JWAF7Y.
Work by P.\v{S}. was supported by the U.S. Department of Energy (DOE), Office of Science, Basic Energy Sciences (BES) under Award Number DE-SC0025265.
\end{acknowledgments}

\bibliography{biblio}

\clearpage 

\onecolumngrid
\setcounter{secnumdepth}{3}
\renewcommand{\thesubsection}{\Roman{subsection}}
\captionsetup[figure]{labelfont={bf},name={Supplementary Figure},labelsep=period}
\setcounter{figure}{0}

\section*{Supplementary Materials}

\subsection{Polymorph growth}
We perform additional simulations to test the growth of a crystalline slab of each polymorph through the [100] cubic plane. Our purpose is to explicitly verify if one polymorph can grow on a surface of a different polymorph. Simulations are done in the $NPT$ ensemble. As illustrated in Supplementary Figure~\ref{fig:coex_c11_in} for the DC-8 structure, the initial configuration involves placing a crystal slab in direct coexistence with liquid particles. The crystal is oriented such that the plane we want to analyse constitutes the interface with the liquid phase (Supplementary Figure~\ref{fig:coex_c11_plane}). The final snapshot in Supplementary Figure~\ref{fig:coex_c11_fin} shows how the initial crystalline slab expands to occupy the entire box, without the emergence of any other polymorph. Hence we can conclude that the lack of trajectories self-assembling the smallest unit cell size polymorph cannot be attributed to the DC-8 initially nucleating and then changing identity as it grows. Others crystalline planes were not analysed. Nevertheless, by tracking the time evolution of the number of crystalline particles of each polymorphs and by visual inspection of initial configurations of nucleating trajectories, we confirm that each polymorph originates and grows from a nucleus of the same polymorph.

\begin{figure}[H]\centering
\subfloat[]{\includegraphics[width=0.3\textwidth]{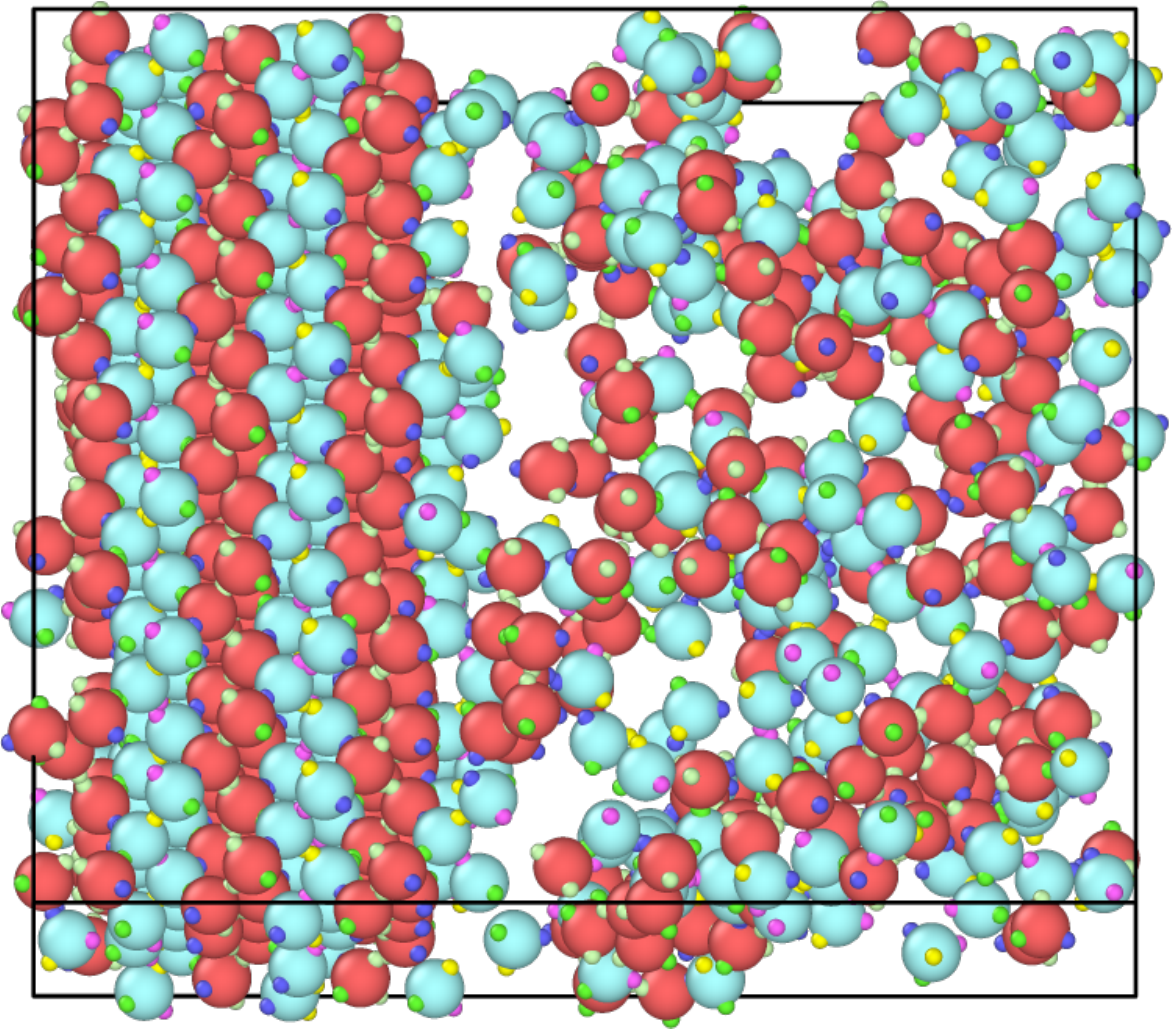}\label{fig:coex_c11_in}}
\hfil
\subfloat[]{\includegraphics[width=0.33\textwidth]{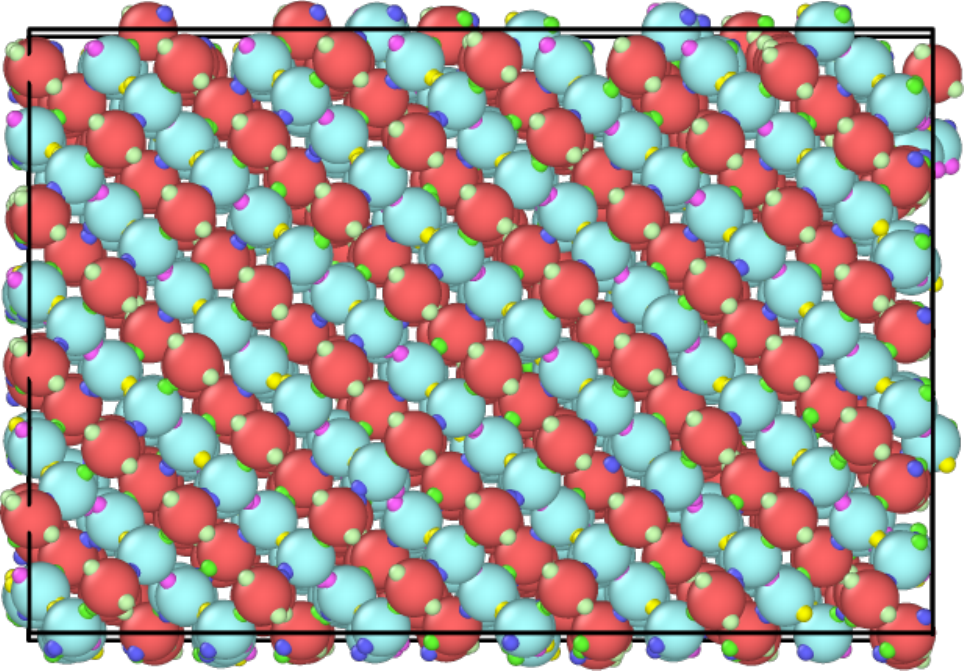}\label{fig:coex_c11_plane}}
\hfil
\subfloat[]{\includegraphics[width=0.22\textwidth]{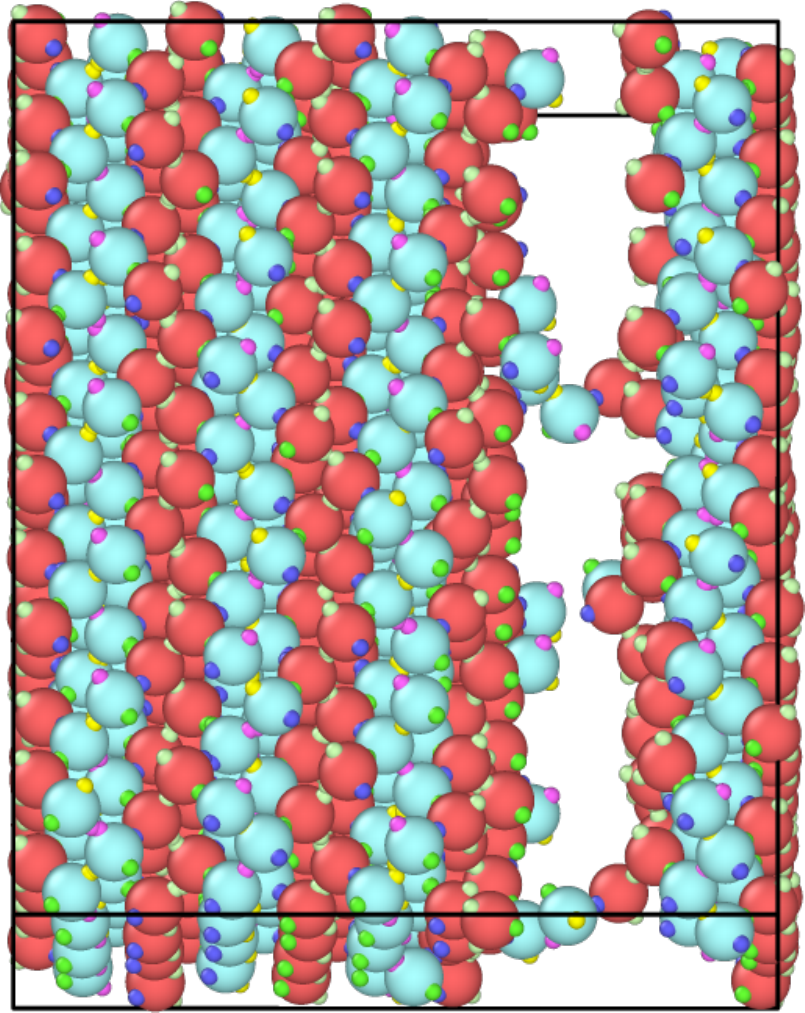}\label{fig:coex_c11_fin}}
 \caption{\textbf{DC-8 growth.} Snapshots of DC-8/liquid coexistence from NPT simulation at T=0.096 and P=0.02. An initial slab of crystal is added to the box with liquid particles (a) with the plane (b) serving as the interface. The crystal rapidly grows eventually spreading throughout the entire box (c). }
\label{fig:c11_growth}
\end{figure}

\subsection{Total coherence order parameter}

In order to classify particles to be in a liquid or crystalline phase we use local bond order analysis. For each particle $i$ the complex vector $q_{lm}(i)$ is defined as: 
\begin{equation}
    q_{lm}(i)=\frac{1}{N_b(i)}\sum_{j=1}^{N_b(i)} Y_{lm}(\mathbf{r_{ij}})
\end{equation}

\noindent where $j$ runs over the $N_b(i)$ nearest neighbours of particle $i$, $Y_{lm}$ are spherical harmonics, $\mathbf{r_{ij}}$ is the vector from particle $i$ to particle $j$, $m$ is an integer whose values range from $m=-l$ to $m=l$ and $l$ is a free parameter (here set to $12$). Two particles are connected if the coherence $S_{ij}$ is larger than a certain threshold $t_{ij}$. $S_{ij}$ is defined as the scalar product $\sum_{m=-12}^{12} Q_{12m}(i)Q_{12m}^{*}(j)$, where

\begin{equation}
    Q_{lm}=\frac{1}{N_b(i)}\sum_{k=0}^{N_b(i)} q_{lm}(k)
\end{equation}

\noindent With $Q_{lm}$ we take into account also the contribution from the second shell around particle $i$ and not only the first shell as one would do by simply using $q_{lm}$. We set $t_{ij}=0.75$, $N_b(i)=16$ and $2.5$ as the maximum distance to consider two particles as neighbours. Usually a particle $i$ is labeled crystalline if its number of connections with its neighbours exceeds a certain value. In this work, instead, we evaluate if the total coherence $\sum_{j} S_{ij}$ is larger than $14.2$. This threshold is set by analysing the total coherence distribution of both pure liquid and thermalised perfect crystal configurations as well as mixed configurations taken from numerical simulations. It is worth noting that this order parameter can effectively label particles as liquid or crystalline, but is not able to distinguish between the three polymorphs since their total coherence histograms perfectly overlap. Other methods based on the Steinhardt order parameter~\cite{steinhardt1983bond}, such as the one introduced by Frenkel~\cite{rein1996numerical} or the one using a two dimensional analysis, whether averaged~\cite{lechner2008accurate} or not~\cite{desgranges2008crystallization}, are equally ineffective. Indeed, all three structures are cubic diamond crystals with the same lattice positions and the same tetrahedral bond geometry. They only differ in the species located at each lattice point. Hence, we develop a system specific order parameter that incorporates the species information. By looking at the three perfect crystals we observe that it is possible to classify them as DC-8, DC-16 and DC-24 with a criterion that only relies on the distance between a particle and its third neighbour of the same species. By counting how many times each distance occurs we can identify the crystal. However this method does not work with thermalised crystals whose distances are not sharply defined. Moreover it is a global criterion that does not provide a label to each particle. If we are interested in classify each particle we have to extract information from the interaction matrix too; in this way we can determine neighbours through the Kern-Frenkel potential definition and keep track of the orientation of each patchy particle. Indeed a particle can be labeled as DC-8, DC-16 or DC-24 by looking at the relative orientation between its first and second neighbours of the same species. In details, we measure the angles $\hat{a}_1$ and $\hat{a}_2$ for the first neighbours and the angle $\hat{a}_3$ for the second neighbours, as shown in Supplementary Figure~\ref{fig:angle1} taking as an example the first species of the DC-8 polymorph. 

\begin{figure}[H]\centering
\subfloat[]{\raisebox{0.5cm}{\includegraphics[width=0.3\textwidth]{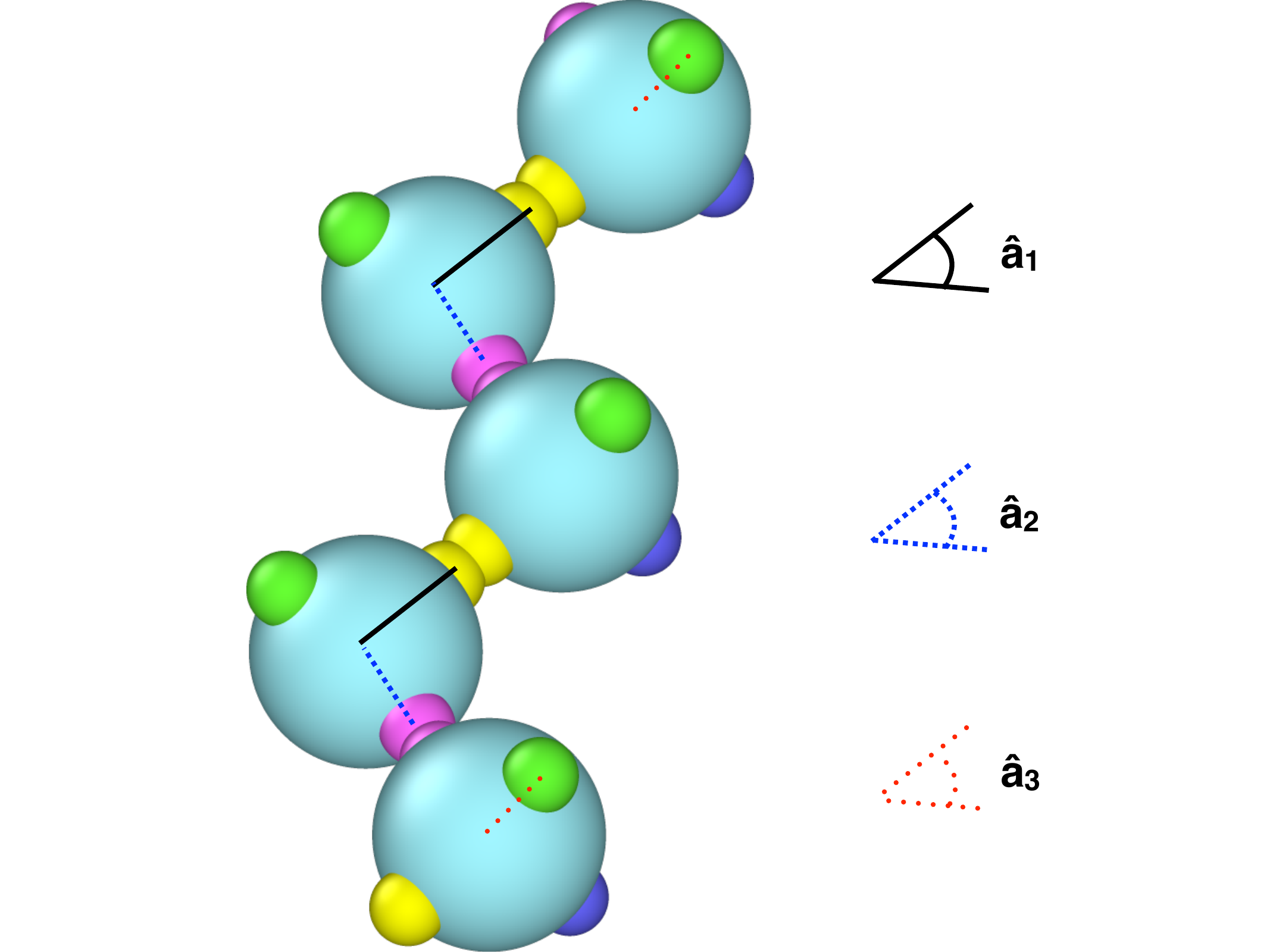}}\label{fig:angle1}}
\subfloat[]{\includegraphics[width=0.33\textwidth]{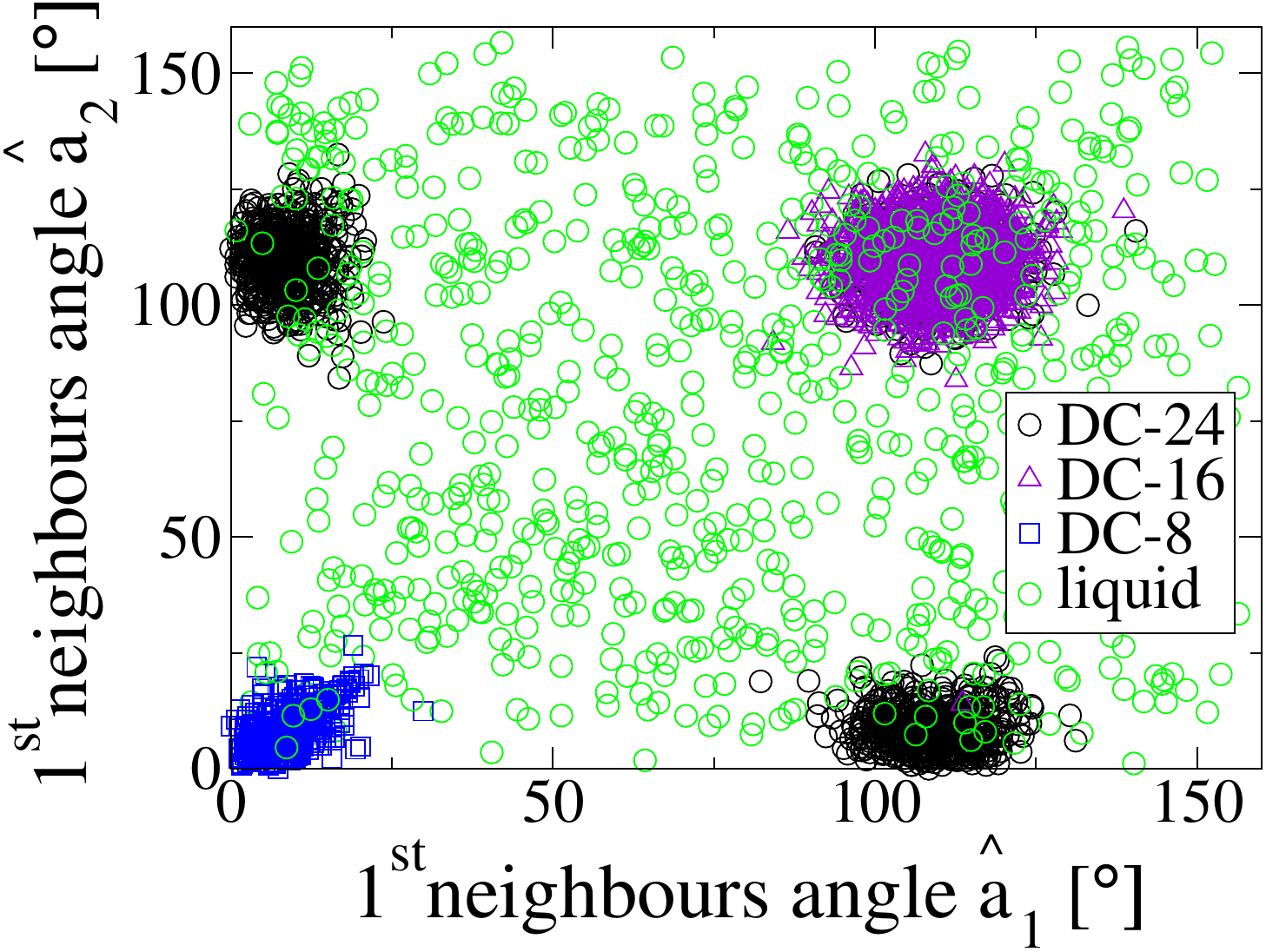}\label{fig:angle2}}
\hspace{2ex}
\subfloat[]{\includegraphics[width=0.33\textwidth]{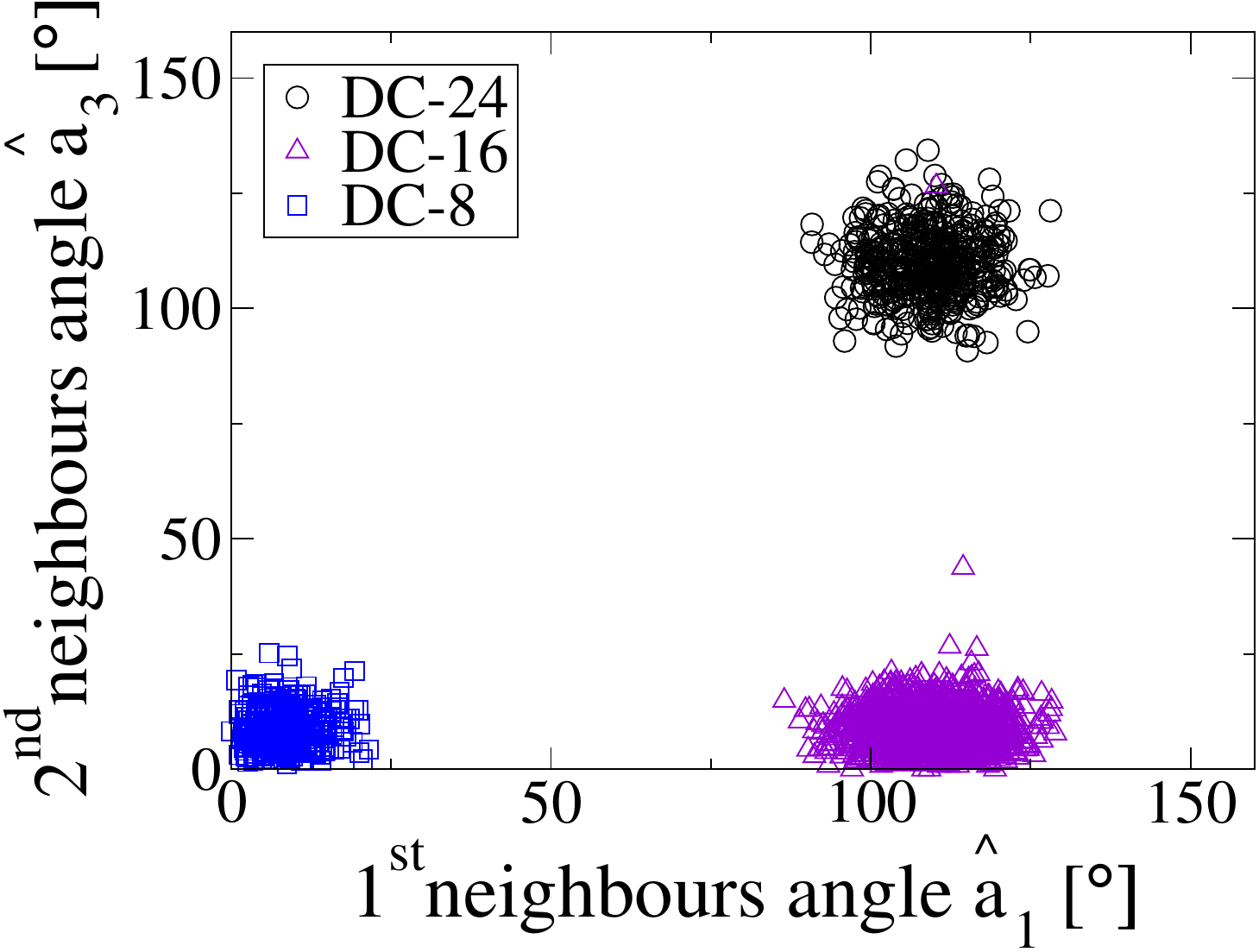}\label{fig:angle3}}
 \caption{\textbf{Orientational order parameter to distinguish between the three polymorphs.} We are able to assign a polymorph label to each particle by analysing the orientation of first (b) and second (c) neighbours of the same species: angle $\hat{a}_1$ and $\hat{a}_2$ of first neighbours and $\hat{a}_3$ of second neighbours are evaluated. (a) DC-8 arrangement of particles of the first species that serves as an example to illustrate the three analysed angles. For perfect crystals, if the first neighbours have the same orientation ($\hat{a}_1$=$\hat{a}_2$=0), we deal with a DC-8, whereas if $\hat{a}_1$=0 and $\hat{a}_2$=109 (or $\hat{a}_1$=109 and $\hat{a}_2$=0) it is a DC-24. If $\hat{a}_1$=$\hat{a}_2$=109 it could be either a DC-24 and a DC-16 and we need second neighbours angle $\hat{a}_3$: if $\hat{a}_3$=0 it is a DC-16 while if $\hat{a}_3$=109 it is a DC-24. In simulations we do not have perfect crystals so angles widen around the previous values as showed in (b) and (c) using black circles for DC-24, purple triangles for DC-16, blue squares for DC-8 and green circles for liquid. 
 }
\label{fig:pp_or}
\end{figure}

\noindent In the perfect DC-8 crystal, particles have the same orientation, i.e. $\hat{a}_1=\hat{a}_2=\hat{a}_3=0$. Conversely, the perfect DC-16 is characterized by $\hat{a}_1=\hat{a}_2=109$ and the perfect DC-24 either by $\hat{a}_1=0$ and $\hat{a}_2=109$ (or vice versa by $\hat{a}_1=109$ and $\hat{a}_2=0$) or by $\hat{a}_1=\hat{a}_2=109$. This is why we need to go to second neighbours; we can distinguish the DC-24 structure from the DC-16 one since for the first $\hat{a}_3=109$, whereas for the second $\hat{a}_3=0$. In simulations we do not have perfect crystals so the angles are not sharply defined, and their distribution spreads out around the previous values. Measured  $\hat{a}_1$, $\hat{a}_2$ (and $\hat{a}_1$, $\hat{a}_3$) angles in simulations are shown in Supplementary Figure~\ref{fig:angle2} (and Supplementary Figure~\ref{fig:angle3}). Purple triangles are used to mark DC-16 angles and black circles represent the DC-24 ones. Data come from nucleating trajectories ending in an extended crystal.

\subsection{Patchy Particles}
Patchy particles are a convenient model to describe short-ranged anisotropic interactions~\cite{kern2003fluid}. They are spherical hard core colloids of diameter $\sigma$ decorated on their surfaces by attractive sites termed patches. Patches are characterized by a depth $\delta$ and a wideness $\theta_{max}$ and the Kern-Frenkel potential we employ to define their interactions relies only on these parameters. Its expression is reported in eq.3:

\begin{equation}
	\label{eqn:KF}
	V(\mathbf r_{ij},\hat{\mathbf r}_{\alpha,i},\hat{\mathbf r}_{\beta,j})= V_{SW}(r_{ij})F(\mathbf r	_{ij},\hat{\mathbf r}_{\alpha,i},\hat{\mathbf r}_{\beta,j})
\end{equation}

\noindent where $\hat{\mathbf r}_{\alpha,i}$ ($\hat{\mathbf r}_{\beta,j}$) indicates the position of patch $\alpha$ ($\beta$) of particle $i$ ($j$), 
\noindent and

\begin{equation}
	\label{eqn:f_KF}
	F(\mathbf r_{ij},\hat{\mathbf r}_{\alpha,i},\hat{\mathbf r}_{\beta,j})=
	\begin{cases}
		1 &\text{if}\quad 
		\begin{array}{l}
			\hat{\mathbf r}_{ij} \cdot \hat{\mathbf r}_{\alpha,i} > \cos{(\theta_{max})}\\ 
			\hat{\mathbf r}_{ji} \cdot \hat{\mathbf r}_{\beta,j} > \cos{(\theta_{max})}
		\end{array} \\
		0 &\text{otherwise}
	\end{cases}
\end{equation}

\noindent The Kern-Frenkel potential is a simple square well isotropic potential $V_{SW}$ of depth $\epsilon$ and width $\delta$ modulated by an orientation dependent term $F$ in such a way that two patchy particles are bonded only if the radius to radius vector $\mathbf r_{ij}$ is shorter than $\sigma+\delta$ and if it intersects the volume of both patches involved in the bond. If instead it is less that $\sigma$ the interaction is infinitely repulsive (HS behaviour).

\noindent Different species of patchy particles can differ by the number (the valence), the arrangement, the type, the depth and wideness of the patches. By tuning these values, different phase diagrams are observed~\cite{smallenburg2013liquids, noya2019assembly}. In this work, the two species of patchy particles share the same parameters ($\delta = 0.2$ and $\cos(\theta_{max})=0.98$), the same valence (4) and the same arrangement (tetrahedral) differing only in the type of the patches (different patches have a different bonding partner). 

\subsection{SAT formulation for inverse self-assembly}
The inverse self-assembly goal is to design building units such that they spontaneously self-assemble into a desired structure. Using a single component system is easier from a thermodynamic perspective, but often it is not experimentally feasible since complex interaction potentials ~\cite{dijkstra2021predictive} and/or shapes tuning~\cite{zhang2005self} are required to avoid the appearance of alternative structures. To maintain the pair interaction potential as simple as possible one approach is to use mixtures. However this is challenging because increasing the number of components creates a combinatorial problem: managing all possible solutions and selecting the optimal one. Here we briefly outline our methodology. We choose patchy particles as building units and we describe their interaction through the manageable Kern-Frenkel potential. Firstly, the number and the placement of the patches are defined to match the bond geometry of the target structure; since in this work we analyze the cubic diamond crystal, we employ patchy particles with four patches tetrahedrally arranged. Then, we define the topology of the target structure unit cell i.e. where each particle is located, what its neighbours are and through which patches bonds are established. What is left to determine is how many different types of species and patches are needed and how they should interact in such a way that the target structure is the minimum free energy fully-bonded state. This task is converted into a colouring problem: by deciding that different colours encode different binding properties, the goal is to define how each patch of each species is coloured and which colours are complemetary allowing the formation of a bond. Since it is impossible to proceed by looking directly for all possible combinations, we transform the coloring problem into a Boolean satisfiability one (SAT) that can be solved efficiently through SAT solvers~\cite{een2005minisat}. The mapping is done by defining binary variables and binary clauses. The first ones encode the design problem and the second ones impose constraints that variables need to satisfy. For instance there are colour interaction variables $x_{c_i,c_j}$ that are true (or equal to 1) if  colour $c_i$ is complementary to color $c_j$. Patch coloring and unit cell placement variables are required too. All the necessary variables and clauses are presented in ref~\cite{russo2022sat}. It is worth noting that boolean clauses are expressions connecting, through the logic operators OR, two ore more binary variables (or their negation with the NOT operator).  The different clauses are linked with AND operators because a solution to the design problem is found if it exists a set of boolean variable values ensuring all clauses are simultaneously satisfied. However the solution can be appropriate also for other structures (as the hexagonal diamond in our example). For this reason the SAT algorithm must be applied iteratively by adding new clauses excluding specific solutions that satisfy also alternative topologies. If the competing structures are known it is not necessary to test each solution by performing numerical simulations. Indeed SAT can be formulated to check if there is a lattice placement for all or a subset of species that is compatible with a particular topology by keeping fixed the patch coloring and the colour interactions variables. 

\noindent The SAT framework is an effective, general, and versatile approach to determine designs that give rise to the desired assembly without defects and stacking faults. Importantly, by combining it with patchy particles model we outline a way to experimentally realize the proposed designs that is not system specific. Moreover, we underline that it is always possible to add additional clauses to enforce specific properties. The only drawback is that it may require more species and/or colours. For example, an additional requirement that we have already analyzed is to design mixtures with an azeotrope at a desired concentration; advantages are twofold: (i) self-assembly is optimized and (ii) the thermodynamic difficulty of dealing with a mixture is overcome~\cite{beneduce2023engineering}. 

\noindent In this paper, we take advantage of another ability of the SAT-assembly framework: the possibility to list all the possible ways to fill the lattice positions of the target unit cell by particles belonging to a selected design. We can indeed insert additional clauses fixing the patch coloring and the colors interactions variable. In this way the SAT solutions represent all the different possible ways these fixed design patchy particles can be arranged in the target unit cell, i.e. all the isotypes. This is a great accomplishment since we can automatically identify all the existing polymorphs of a target lattice of a certain size, regardless of their nucleation ability. For the cubic diamond crystal of 48 particles analyzed in this work, SAT uncovers 64 possible arrangements of patchy particles belonging to the N2c8 design. As shown in Figure 2 in the paper, they can be grouped into three periodically repeated patterns that we name DC-8, DC-16, DC-24 to account for their different unit cell sizes.

\subsection{Umbrella Sampling and Successive Umbrella Sampling}
Nucleation phenomena are activated processes, i.e. in order to happen a free energy barrier must be overcome. We do not aim at fully reconstruct the free energy landscape, but rather we want to know the free energy cost $F(n)$ of forming a nucleus of size $n$ in a metastable phase. We can obtain it from the probability $P(n)$ that the system is in a configuration where a crystalline cluster of size $n$ is formed. Indeed:

\begin{equation}
    F(n)=-k_BT \ log P(n)
\end{equation}

\noindent Conceptually, the probability distribution along the reaction coordinate $n$ is a quantity easily measurable in Monte Carlo simulations. However, when states of interest are rarely accessed, the statistical accuracy is poor and direct sampling is ineffective. This is exactly the case of nucleation phenomena. Indeed, in this scenario the system is stuck in a relative free energy minimum corresponding to the metastable phase (liquid), and eventually, when a rare fluctuation great enough to overcome the barrier (considerably larger than $k_BT$) occurs, it reaches the  absolute minimum and we start sampling configurations in the stable phase (solid). This means that with direct sampling we effectively probe only the configuration space around minima. However, to construct the nucleation free energy barrier, we need to sample the configuration space where crystalline clusters from small sizes to the critical one are formed, i.e we aim to sample configurations along and on top of the barrier. We can do it by modifying the Hamiltonian of the system when computing the acceptance probability of a move: an additional term, named biasing potential $V_{bias}(n)$, is added in order to ensure that the system visits regions that it might not otherwise explore sufficiently. The biasing potential depends only on the reaction coordinate $n$ and $V_{bias}(n)=-F(n)$ would be the ideal choice since it guarantees a uniform sampling in the whole reaction coordinate interval. However $F(n)$ is not known and it is exactly what we want to compute. A solution is provided by the Umbrella Sampling (US) method: the range of interest of the reaction coordinate is divided into small windows and a potential that constraints the system to explore regions close to the desired value of $n$ is applied in each of them. The name derives from the biasing potential effect to bridge energetically separated regions. Often an harmonic potential is chosen such that:

\begin{equation}
    H=H_{system}+V_{bias}(n) \text{ with } V_{bias}(n)=k(n-n_0)
\end{equation}

\noindent The system is forced to explore the phase space regions around $n_0$ with fluctuations depending on the value of $k$. By using a bias potential, we measure a bias distribution $P_{bias}(n)$ in each window, but we can easily retrieve the original unbiased one:

\begin{equation}
P(n)=P_{bias}(n)e^{\beta V_{bias}(n)}
\end{equation}

\noindent The overall $P(n)$ is computed by combining the results from the different windows, therefore it is necessary that single windows distributions overlap with the neighbour ones. This means that the strength of the bias $k$ must be carefully calibrated and may change for different windows (in our case, a constant value
 of $k=0.005$ was sufficient). For instance a high $k$ is required if the free energy is steep. However we do not know \textit{a priori} the slope of the free energy. To minimize the problem of selecting a bias potential for each window that ensures proper sampling, Virnau and Muller proposed the Successive Umbrella Sampling technique (SUS)~\cite{virnau2004calculation}, a kind of US in the limit of small windows size. In each window $i$ the system is constrained to explore values of the reaction coordinate $n$ belonging to a small interval of size $\omega$ ($n \in [i\omega,(i+1)\omega])$; Monte Carlo simulations are run in the grand-canonical ensemble where all moves bringing the system outside the window ranges are rejected. The windows size $\omega$ is so small, usually equal to $2$ or $3$, that the use of biasing potential is not necessary. Free energy difference between a state with reaction coordinate of $n$ and the reference one $n=0$ can be computed through the $P(n)/P(0)$ ratio. Here, this ratio is reconstructed by the product of the ratios between the boundary histograms of each window.

\begin{equation}
    \frac{P(n)}{P(0)}=\frac{H_{0r}}{H_{0l}} \frac{H_{1r}}{H_{1l}} \cdots \frac{H_{nr}}{H_{nl}}
\end{equation}

\noindent Indeed, during the simulation, for each independent window $i$, the number of times each configuration $n$ is visited is saved into a histogram $H_i[n]$. It is important to underline that, in order to satisfy the detailed balance, the left boundary histogram $H_{il}=H_i[i\omega]$ (and the right one $H_{ir}=H_i[(i+1)\omega]$) must be incremented by one even when we reject the moves overcoming the lower (or the upper) limit of the interval. By computing the non boundary histograms it is possible to define an adaptive weight if a biasing potential is still used.

\noindent Both US and SUS are powerful techniques that enables to compute free energy projection along a reaction coordinate without knowing the free energy behaviour.

\subsection{Liquid-solid interfacial tension with BSUS}

\begin{figure}[H]
\subfloat[]{\includegraphics[width=0.48\textwidth]{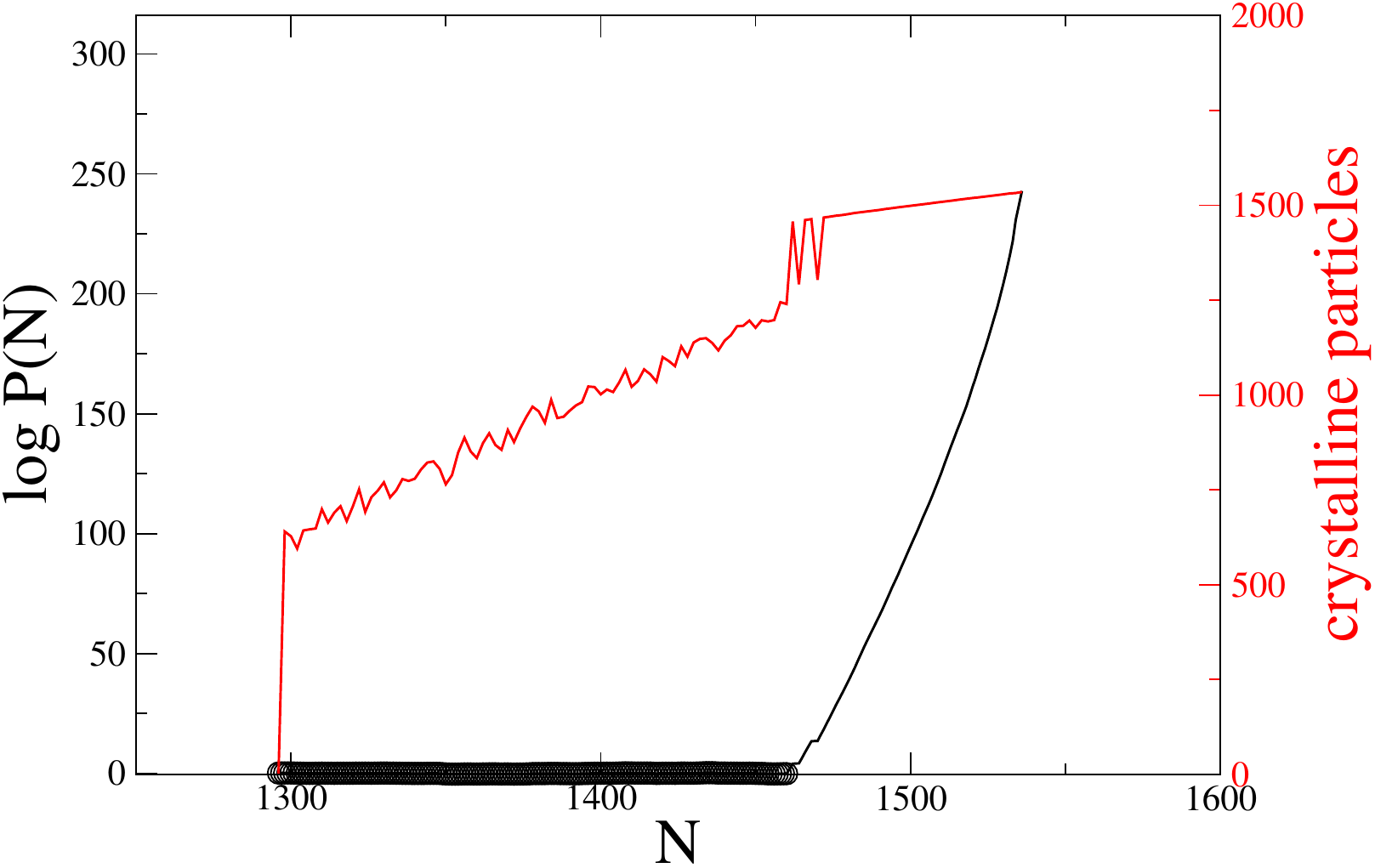}\label{fig:pn_op_bsus}}
\hfil
\subfloat[]{\includegraphics[width=0.4\textwidth]{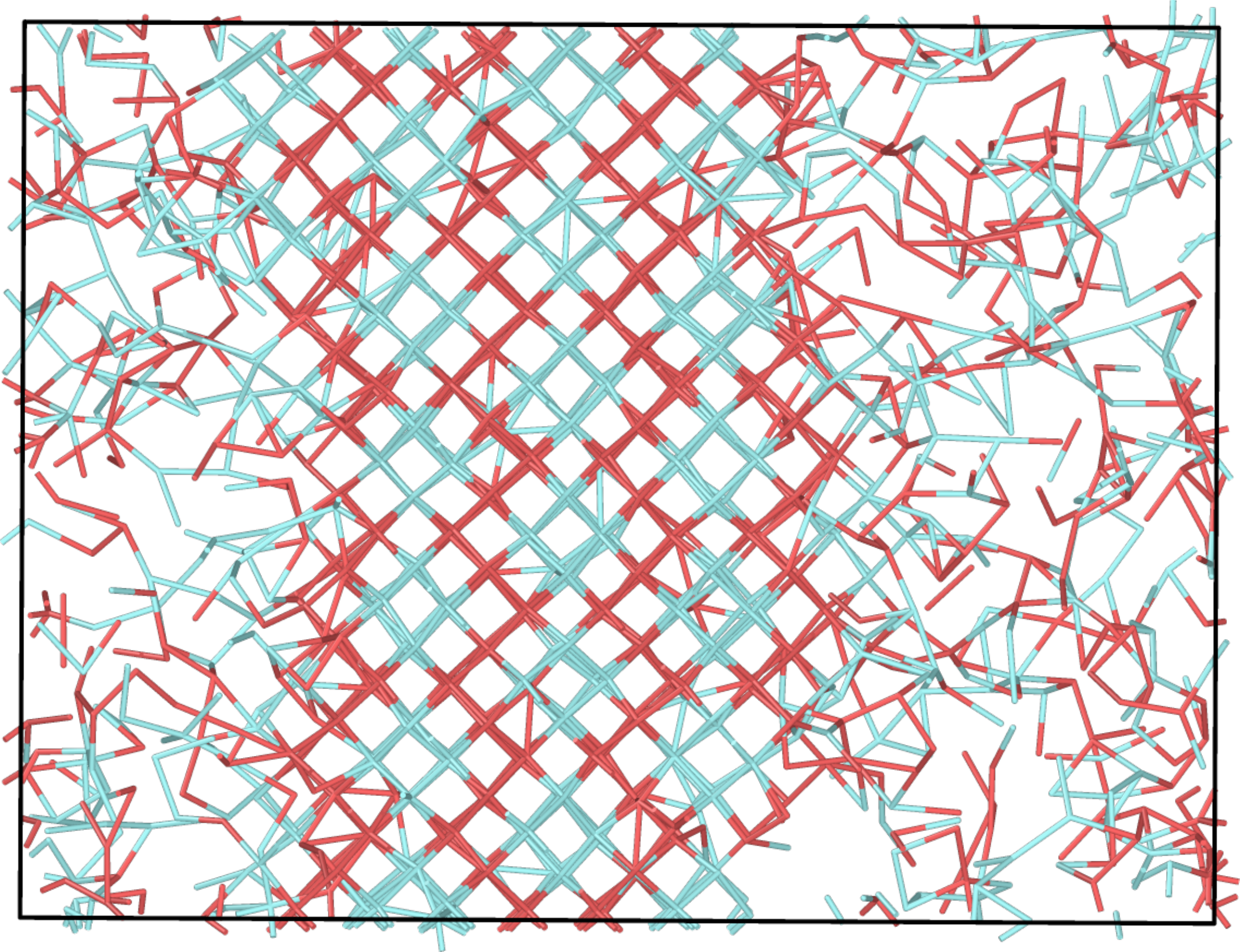}\label{fig:coex_bsus}}
 \caption{\textbf{Surface tension computation.} (a) Not normalized probability distribution function $logP(N)$ (black line) from BSUS simulations of DC-16 cubic diamond crystal and liquid coexistence at $T=0.133$ and $z=2.25$. The distribution has been reweighted to have a flat profile at coexistence, as highlighted by black circles. The number of crystalline particles (red line) emphasizes the coexistence and the crystalline regions. 
 (b) Snapshot of a configuration from BSUS window of 1350 particles showing the two coexisting phases; only bonds are represented.}
\label{fig:bsus_plot}
\end{figure}

In order to explicitly verify that the interfacial properties of the three polymorphs are the same, we estimate their surface tensions. We apply the bias successive umbrella sampling technique (BSUS) testing this approach for the first time on liquid-solid phases of a patchy particles system.
Monte Carlo simulations are run in parallel in distinct but overlapping  density windows in the grand-canonical ensemble $(\mu V T)$; the first window consists of a perfect crystal box of $N=1536$ particles and the successive ones are initialized by removing particles from the crystal, two by two until $N = 1290$, in the direction perpendicular to the selected interface. We choose a window size of three particles. In each window the number of times each possible state ($\{N_{i-1}, N_i, N_{i+1}\}$) is visited is counted in order to eventually reconstruct the probability distribution $log(P(N))$. Simulations are performed over a specified number of Monte Carlo sweeps. Each MC sweep represents a single iteration of the simulation, during which multiple moves are attempted across the system. The density at which the crystal is prepared is the coexisting one $\rho_{coex}=0.52$, as evaluated by direct coexistence simulations in $NPT$ ensemble at $T=0.133$ and $P=0.2$. The activity $z$ is fixed at $2.4$ and the number of crystalline particles in each window is also calculated. Its behaviour through the $124$ windows is shown in red in Supplementary Figure~\ref{fig:pn_op_bsus}: the coexistence and the solid regions are easily identified. In the coexistence region the number of crystalline particles grows linearly, while at $N=1300$ it straightly goes to zero indicating the suddenly melt of the crystal due to interface fluctuations. A snapshot of a configuration at coexistence is  reported in Supplementary Figure~\ref{fig:coex_bsus}. Identifying the coexistence region allows us to reweight the probability distribution $log(P(N))$ in order to obtain the equilibrium one $log(P_{coex}(N))$ that is plotted in  black in Supplementary Figure~\ref{fig:pn_op_bsus}. Indeed the equilibrium distribution must be flat in the coexisting region since, at coexistence, there is no free energy cost in moving the interface along the direction perpendicular to it. Since $P_{coex}(N)=P(N)e^{\beta((\mu_{coex}-\mu)N}$, in a logarithmic y-axis plot we can straightforwardly determine the coexistence activity $z_{coex}=e^{\beta \mu_{coex}}$ by linear fitting the coexistence region of $log(P(N))$ vs. $N$. Then we can scale the distribution such that its slope is effectively zero at coexistence. Having now the distribution at equilibrium we compute the interfacial tension by measuring the height of the crystal peak and dividing it by two times the area of the interface. Indeed, as visible in Supplementary Figure~\ref{fig:coex_bsus}, the planar interfaces are two since periodic boundary conditions are active.  This procedure is repeated few times for each polymorph and the resulting average values of interfacial tension are reported in Figure 2 of the paper. We can assert that they are all the same within an error of $3\%$, thus confirming that the three polymorphs can be considered equal in their interfacial properties. 

\subsection{Check against the existence of unfavorable surface sites}

\noindent In this section, we aim to verify that the three polymorphs do not differ in the density of attachment on surface sites. In other words, we want to rule out the possibility that differences in the nucleation free energy barriers can be attributed to the presence of favorable or unfavorable bonding sites on the surface of the three polymorphs. 

\noindent To achieve this, we ran additional Monte Carlo simulations in the canonical ensemble with $1000$ particles. In particular, we prepared the system by inserting in the liquid phase a crystalline supercritical seed of size $100$ (Supplementary Figure 4a).  
To prevent the nucleus from growing, the state point is chosen to be unstable for the crystal phase ($T=0.16, \rho=0.35$), while to prevent it from disappearing, the particles of the seed are kept fixed during the simulations. We evaluated the potential energy difference $(U-U_0)$ of surface particles, relative to the configuration where the nucleus is in vacuum (Supplementary Figure 4b). In this way, we measure the new bonds formed between surface particles and melt ones (Supplementary Figure 4c). We underline that $U_0$ and the patches orientation are the same for all three frozen nuclei. 

\begin{figure}[t]
    \centering
    \subfloat[]{\includegraphics[width=0.3\textwidth]{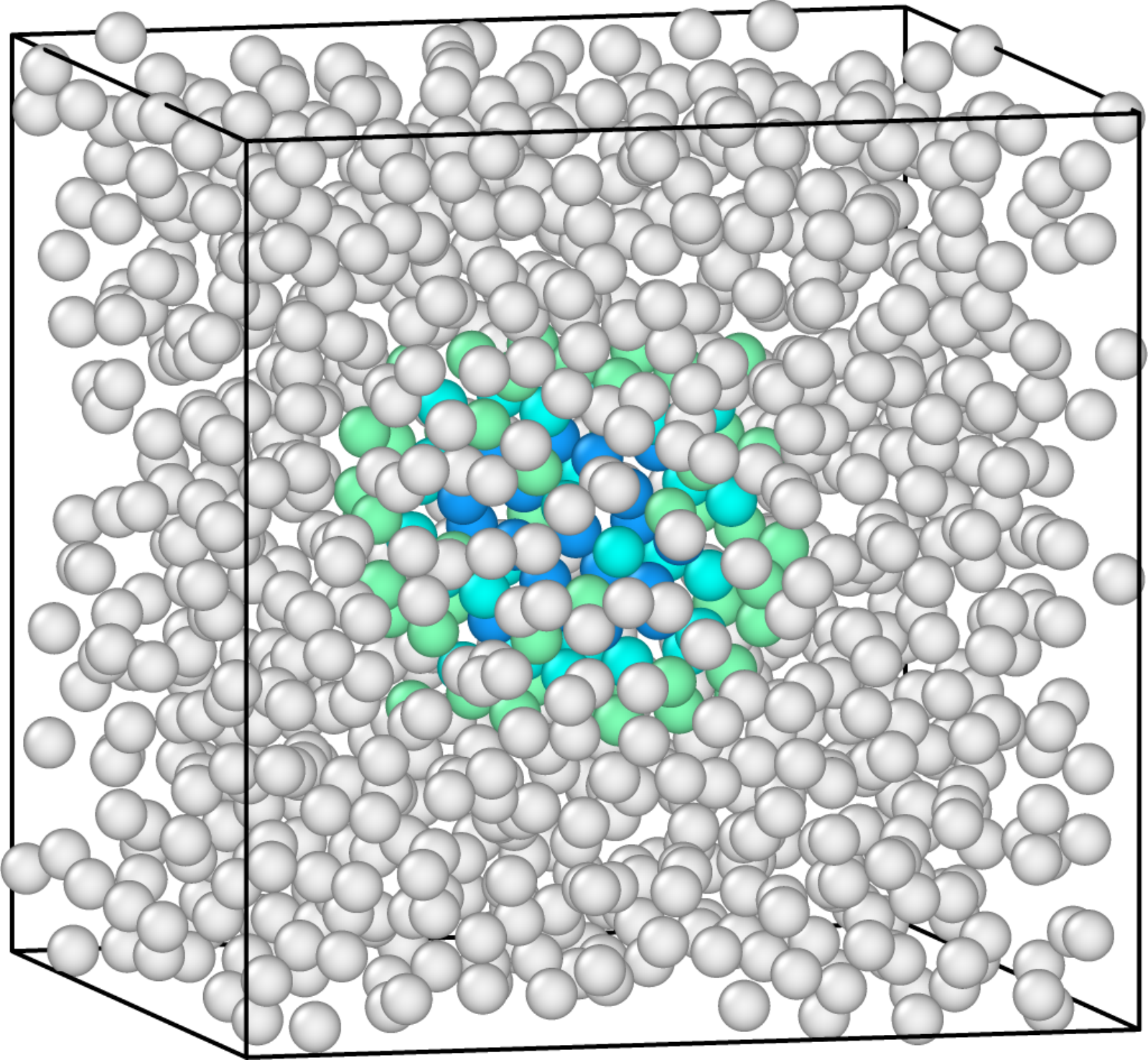}\label{1a}}
    \hfil
    \subfloat[]{\includegraphics[width=0.268\textwidth]{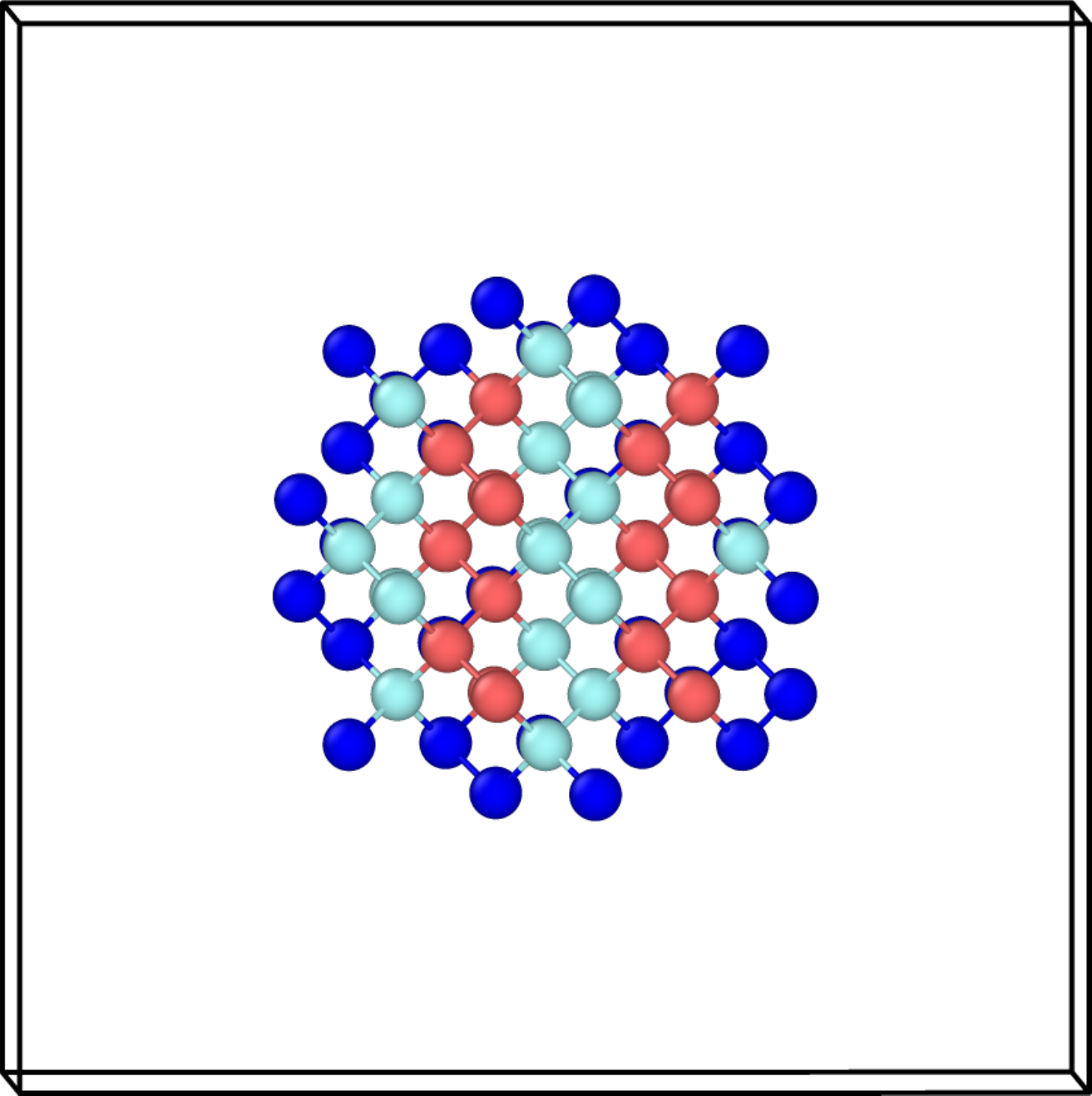}\label{1b}}
    \hfil
    \subfloat[]{\includegraphics[width=0.272\textwidth]{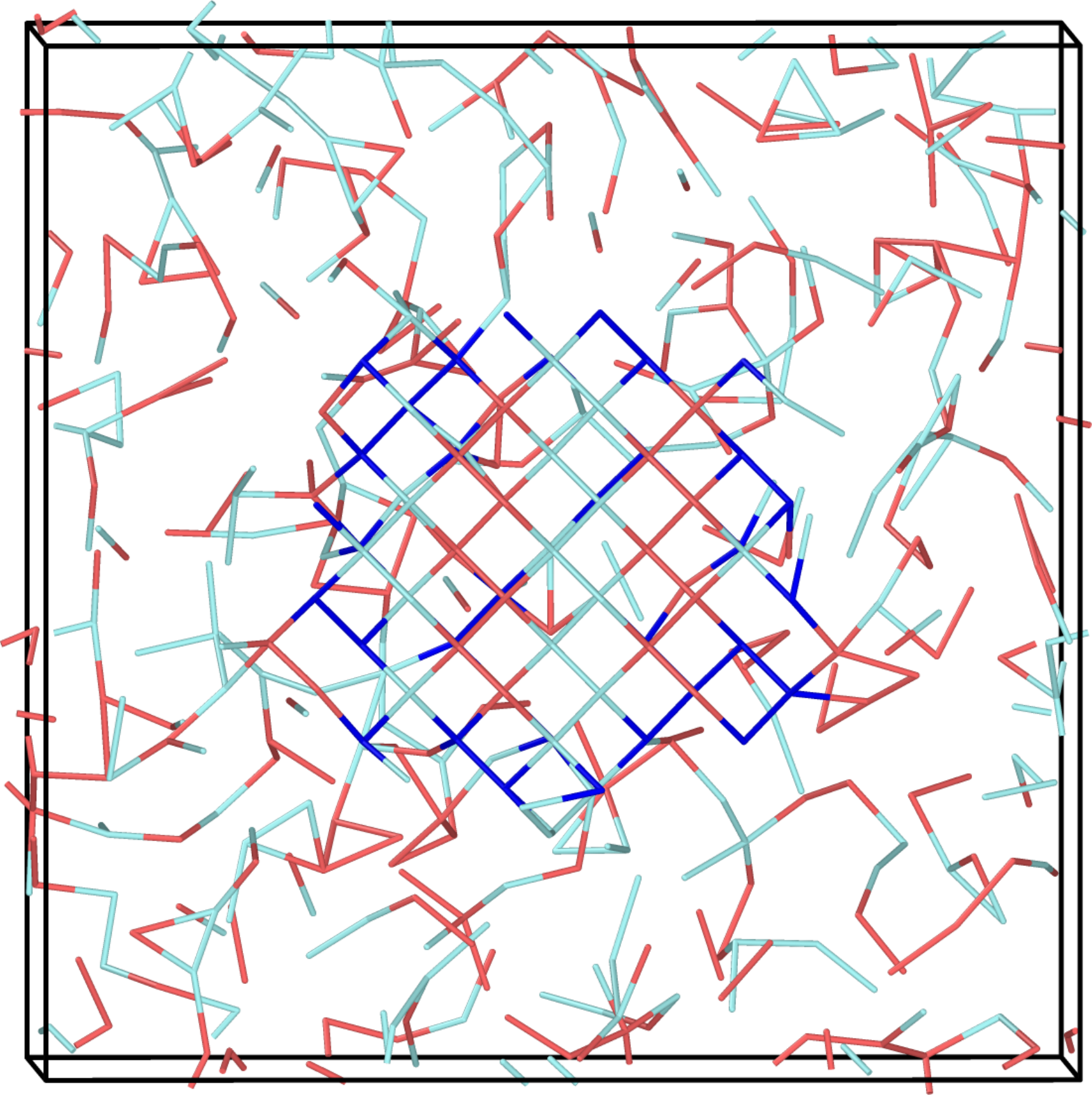}\label{1c}}
   \caption{\textbf{Initial and final snapshots with a frozen DC-8 crystalline seed.} Snapshot of a typical initial configuration where a crystalline seed of $100$ (colored) particles is surrounded by $1000$ liquid (white) particles. (b) The nucleus is frozen and the particles colored blue are surface particles whose energy in vacuum is $U_0$. During the simulation, these surface particles can establish bonds with melt particles (c). This allow us to evaluate how many new nucleus-melt bonds are formed in each crystalline polymorph and determine if there is a structure where bond formation is more favorable. }
\label{fig:snap}
\end{figure}

\noindent In Supplementary Figure 5 we plot, for each polymorph, the average surface-melt bonding probability. It is obtained by normalising the $(U-U_0)$ difference with respect to the maximum number of bonds that surface particles can establish with the melt. Values are obtained by averaging over different configurations and trajectories and they appear to be the same for the three structures. Hence, each nucleus has the same ability to make bonds with melt particles. 

\begin{figure}[H]
    \centering
    \includegraphics[width=0.52\textwidth]{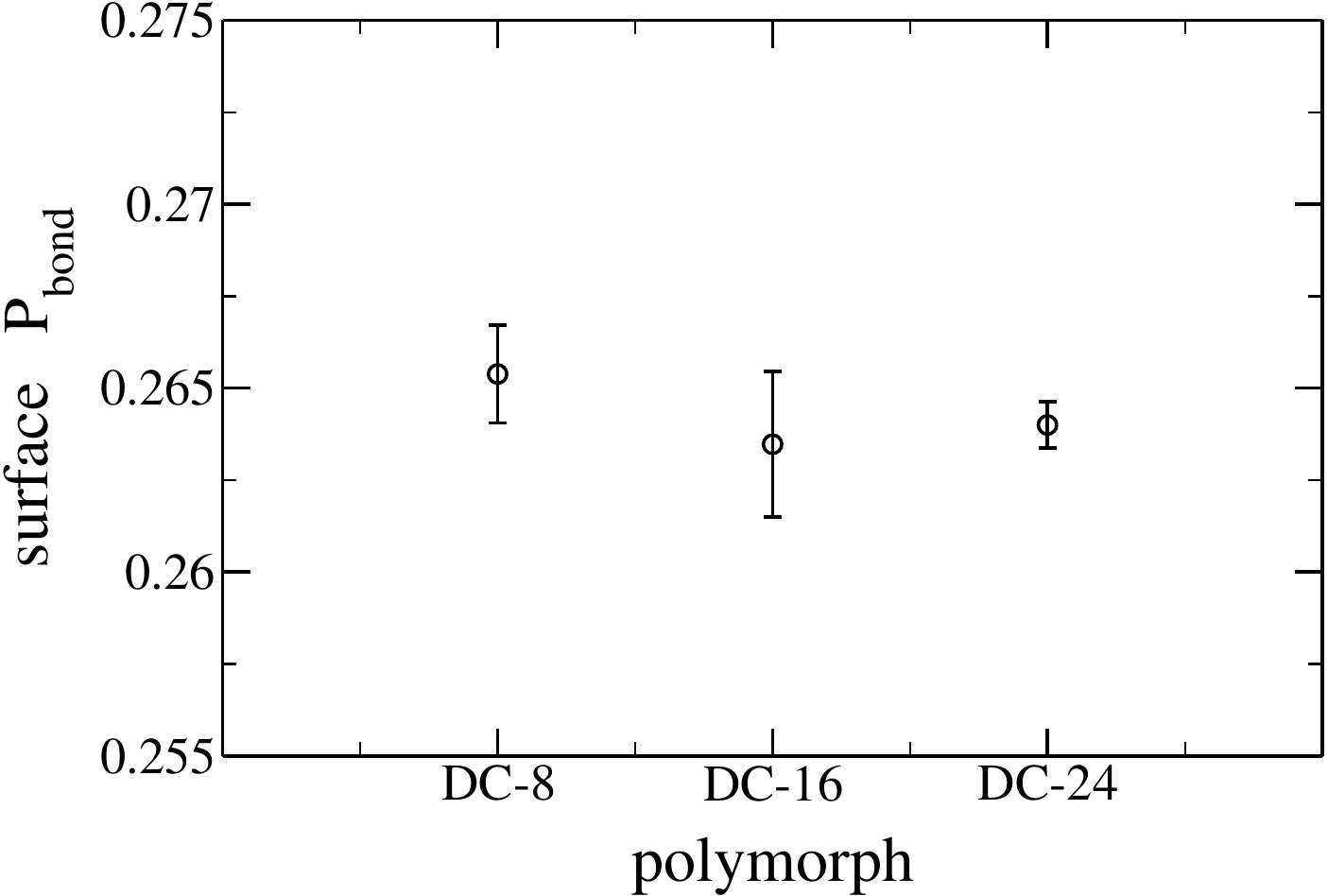}
     \caption{\textbf{Surface-melt particles bonds formation.} The average surface-melt bonding probability is the same (within the error) for the three polymorphs. This indicates that the surface particles of the three nucleus share the same bonding ability.}
\label{fig:deltaE}
\end{figure}

\noindent In conclusion, the differences in the heights of the nucleation barriers cannot be ascribed to the presence of unfavorable surface particles that would lead to an entropic penalty in the less frequent structures. This provides additional evidence that the three polymorphs share the same surface tension. Hence they are completely degenerate and, according to CNT predictions, should not display any differences in their nucleation frequency.

\subsection{The DC-8 case}

\noindent This section is devoted to show that the difference in the nucleation barrier heights of the DC-8 and DC-16 polymorphs can still be traced back to the liquid structure by looking at the distribution of the angle $\alpha_0$. It defines the relative orientation between first neighbours of the same species and it is illustrated in the inset of Supplementary Figure 6b for patchy particles of the first species belonging to the DC-24 structure.

\noindent As discussed in the paper and shown in Figure 4 therein, the DC-8 and DC-16 crystals share common $\alpha$ values that do not align with the typical melt values. However, by plotting the radial profile of $\alpha_0$ angle for the three polymorphs (Supplementary Figure 6b), we observe that their bulk values are different. In particular, the DC-8 angle is the farthest from the melt values. This explains why the DC-8 polymorph has the highest nucleation barrier and the DC-24 polymorph the lowest one. Moreover, the similarity between the DC-16 and DC-24 bulk $\alpha_0$ angles highlights the need to extend the analysis to second neighbours to fully understand the distinct nucleation behaviours of the polymorphs.

\noindent In Supplementary Figure 6a we also show the crystalline profiles for all three polymorphs through their average radial distribution of the total coherence. They allow the identification of the crystalline, the interfacial and the melt regions. Points are computed by averaging on different configurations and trajectories from umbrella sampling windows characterised by a crystalline nucleus of 50 particles.

\begin{figure}[H]
    \centering
    \includegraphics[width=0.6\textwidth]{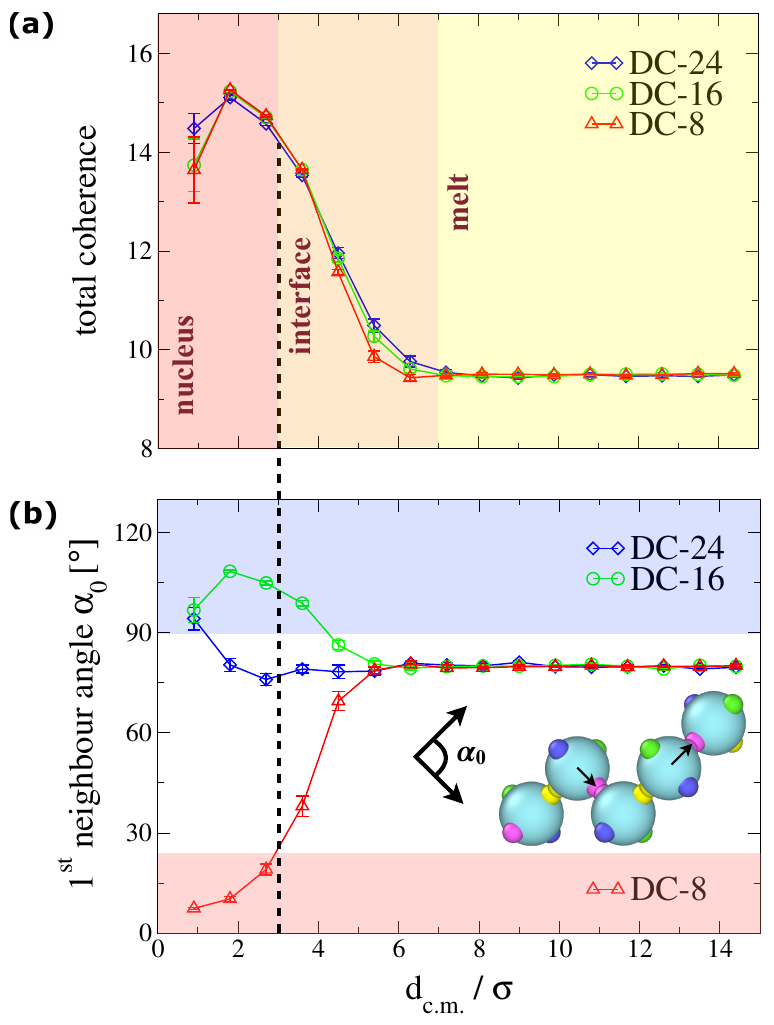}
     \caption{\textbf{First neighbours radial profiles.} Average radial profile of total coherence (a) and $\alpha_0$ angle (b) from the center of mass $d_{c.m.}$ of a 50 particles crystalline nucleus of DC-24 (blue diamonds), DC-16 (green circles) and DC-8 (red triangles) polymorphs. (a) $\alpha_0$ defines the relative orientation between first neighbours of the same species, as illustrated for patchy particles of the first species of the DC-24 polymorph in the inset of figure (b). The coloured bands in (a) helps locating the crystalline, the interfacial and the melt regions. The coloured bands in (b) define the range of $\alpha_0$ values common to the bulk DC-24 and the bulk DC-16 (blue band) polymorphs and the one characteristic of the bulk DC-8 (red band) structure. Red band deviates from the typical values of the melt which are more closely approached by the blue band.}
\label{fig:1angles}
\end{figure}

\end{document}